\documentclass[12pt]{iopart}

\usepackage{graphicx}
\usepackage{iopams}

\begin{document}

\title[ S Longhi, Zitterbewegung of optical pulses ...]{Zitterbewegung of optical pulses in nonlinear frequency conversion}

\author{Stefano Longhi}

\address{Dipartimento di Fisica,
Politecnico di Milano, Piazza L. da Vinci 32, I-20133 Milano, Italy}
\ead{longhi@fisi.polimi.it}

\begin{abstract}
\noindent Pulse walk-off in the process of sum frequency generation
in a nonlinear $\chi^{(2)}$ crystal is shown to be responsible for
pulse jittering which is reminiscent to the Zitterbewegung
(trembling motion) of a relativistic freely moving Dirac particle.
An analytical expression for the pulse center of mass trajectory is
derived in the no-pump-depletion limit, and numerical examples of
Zitterbewegung are presented for sum frequency generation in
periodically-poled lithium niobate. The proposed quantum-optical
analogy indicates that frequency conversion in nonlinear optics
could provide an experimentally accessible simulator of the Dirac
equation.
\end{abstract}


\maketitle

\section{Introduction}
Originally predicted by Schr\"{o}dinger in the study of the Dirac
equation \cite{Schrodinger}, Zitterbewegung (ZB) refers to the
trembling motion of a freely-moving relativistic quantum particle
that arises from the interference between the positive- and
negative-energy parts of the spinor wave function \cite{Grenier}.
For a free electron, the Dirac equation predicts the ZB to have an
extremely small amplitude (of the order of the Compton wavelength
$\simeq 10^{-12} \; {\rm m}$) and an extremely high frequency
($\simeq 10^{21} \; {\rm Hz}$), making such an effect experimentally
inaccessible. In addition, the physical relevance of ZB in
relativistic quantum mechanics is a controversial issue because such
an effect arises in the framework of the single-particle picture of
the Dirac equation, but not in quantum field theory
\cite{Barut,Grobe}. The notion of ZB and resulting formalism,
however, are not peculiar to relativistic quantum dynamics, and
phenomena analogous to ZB, which underly the same mathematical model
of the Dirac equation, have so far predicted in a wide variety of
quantum and even classical physical systems, including among others
semiconductors and quantum wells \cite{ZB1-1,ZB1-2,ZB1-3}, trapped
ions \cite{ZB2}, graphene \cite{ZB3,ZB4-1,ZB4-2,ZB4-3}, cold atoms
\cite{ZB5-1,ZB5-2}, acoustic \cite{acoustic} and photonic
\cite{P1,P2,P3} systems. Simulations of relativistic quantum effects
using experimentally-accessible physical set-ups, in which parameter
tunability allows access to different physical regimes, have seen in
recent years an increasing interest, culminating to the very recent
first experimental observation of a quantum analogue of ZB using a
single trapped ion set to behave as a free relativistic quantum
particle \cite{NatureZB}. In the optical context, the use of
photonic systems to mimic quantum phenomena in the lab has seen a
continuous and increasing interest (see, for instance,
\cite{LonghiLPR} and references therein); in particular, optical
analogues of the relativistic ZB have been recently proposed to
occur in photonic crystals \cite{P1}, metamaterial slabs \cite{P2},
and binary waveguide arrays \cite{P3}. In this work it is shown
theoretically that a classical analogue of ZB can be observed in a
much simpler and well-known set-up of nonlinear optics, namely in
the process of sum frequency generation of light waves in a
nonlinear $\chi^{(2)}$ medium \cite{Boyd} in presence of pulse (or
spatial) walk off. In the nonlinear optics context, optical
three-wave interaction (TWI) in nonlinear $\chi^{(2)}$ media in
presence of temporal (or spatial) walk-off is a well-known process,
which has been widely investigated especially in connection to pulse
compression of ultrashort pulses and to TWI soliton theory (see, for
instance, \cite{S1,S2,S3,S4,S5,S6}). Notably, the nonlinear TWI
equations are solvable by inverse scattering methods \cite{S1}.
However, the ZB phenomenon discussed in this work and the idea of
exploiting nonlinear optics to mimic the Dirac equation have not
been addressed in such previous studies.

\section{Basic model and quantum-optical analogy}
The starting point of our analysis is provided by a standard model
of TWI in a nonlinear quadratic medium describing propagation of
either optical pulses or optical beams at frequencies $\omega_1$,
$\omega_2$ and $\omega_3=\omega_1+\omega_2$ in presence of either
group velocity mismatch or spatial walk-off. For the sake of
definiteness, we will consider here the case of optical pulse
interaction in presence of temporal walk-off (i.e. of group velocity
mismatch), however the results hold also for spatial beam
propagation in a quadratic medium with spatial walk-off provided
that the temporal coordinate is replaced by a transverse spatial
coordinate (see, for instance, \cite{S6}). Assuming that group
velocity dispersion is negligible, in the plane-wave approximation
and assuming perfect phase matching, pulse propagation in the
$\chi^{(2)}$ medium is described by the following set of nonlinear
coupled equations \cite{Boyd,S1,S2,S3,S4,Longhi02}:
\begin{eqnarray}
\left( \frac{\partial}{\partial z}+\frac{1}{v_{g1}}
\frac{\partial}{\partial t} \right) A_1& = & i \rho A_2^*A_3 \\
\left( \frac{\partial}{\partial z}+\frac{1}{v_{g2}}
\frac{\partial}{\partial t} \right) A_2& = & i \rho A_1^*A_3 \\
\left( \frac{\partial}{\partial z}+\frac{1}{v_{g3}}
\frac{\partial}{\partial t} \right) A_3 & = & i \rho^* A_1 A_2
\end{eqnarray}
where $A_l=A_l(z,t)$ ($l=1,2,3$) is the amplitude of the electric
field envelope at the carrier frequencies $\omega_l$, normalized
such that $|A_l|^2$ is the photon flux at frequency $\omega_l$,
$v_{gl}$ is the group velocity in the medium at frequency
$\omega_l$, and $\rho$ is the strength of the nonlinear interaction,
which reads explicitly
\begin{equation}
\rho=\frac{d_{eff}}{c_0} \sqrt{\frac{2 \hbar \omega_1 \omega_2
\omega_3}{\epsilon_0 c_0 n_1 n_2 n_3}},
\end{equation}
where $n_l$ is the refractive index of the medium at frequency
$\omega_l$ ($l=1,2,3)$, $c_0$ is the speed of light in vacuum, and
$d_{eff}=(1/2) \chi^{(2)}_{eff}$ is the effective nonlinear
coefficient. To achieve perfect phase matching, a
quasi-phase-matching (QPM) grating for the nonlinear susceptibility
$\chi^{(2)}$ can be employed; in this case one has (see, for
instance, \cite{Longhi02})
\begin{equation}
d_{eff}=\frac{1}{2} \overline{\chi^{(2)}(z) \exp(i \Delta k z)}
\end{equation}
where $\Delta k=(\omega_3 n_3-\omega_2n_2-\omega_1n_1)/c_0$ is the
phase mismatch of the three waves and the overbar denotes a spatial
average over a few modulation periods of the QPM grating. Equations
(1-3) admit of the following two invariants along the propagation
distance $z$
\begin{equation}
\mathcal{I}_{1,2}= \int_{-\infty}^{\infty} dt (|A_{1,2}|^2+|A_3|^3)
\end{equation}
which correspond to photon flux conservation (Manley-Rowe
invariants) in the frequency conversion process. Solitary waves of
Eqs.(1-3) in the fully nonlinear regime, including trapped
bright-dark-bright solitary waves with locked velocity, have been
investigated in Refs.\cite{S1,S2,S3,S4,S5,S6}. To study the analogue
of ZB in the frequency conversion process, we assume here that at
the input plane $z=0$ the nonlinear crystal is excited by a strong
and nearly continuous-wave pump field at frequency $\omega_1$, and
by a weak and short signal pulse at frequency $\omega_2$ and
temporal profile $g(t)$. Under such assumptions, the invariance of
$\mathcal{I}_{1,2}$ implies that the pump wave remains nearly
undepleted along the propagation distance, and Eqs.(1-3) reduce to
the following two linear coupled-field equations describing
sum-frequency generation in the undepleted regime
\begin{eqnarray}
\left( \frac{\partial}{\partial z}+\frac{1}{v_{g2}}
\frac{\partial}{\partial t} \right) A_2& = & -i \kappa A_3 \\
\left( \frac{\partial}{\partial z}+\frac{1}{v_{g3}}
\frac{\partial}{\partial t} \right) A_3 & = & -i \kappa A_2
\end{eqnarray}
where we have set $\kappa \equiv -\rho A_1^*=\rho {\sqrt {I_1/ \hbar
\omega_1}}$ and $I_1$ is the intensity of the pump field. Without
loss of generality, $\kappa$ can be assumed to be real-valued and
positive. Note that, in the absence of group velocity mismatch for
the signal and second-harmonic waves at frequencies $\omega_2$ and
$\omega_3$, i.e. for $v_{g2}=v_{g3}$, the solution to Eqs.(7) and
(8) is analogous to the one for stationary fields \cite{Boyd}, which
shows a well-known oscillatory power transfer, along the propagation
distance $z$, between the two fields with spatial period
$\pi/\kappa$; namely one has
\begin{eqnarray}
A_2(z,t) & = & g \left( t-\frac{z}{v_g}\right) \cos(\kappa z) \\
A_3(z,t) & = & -i g \left( t-\frac{z}{v_g}\right) \sin(\kappa z)
\end{eqnarray}
where $v_g=v_{g2}=v_{g3}$. Note that, in spite of the oscillatory
power transfer in the frequency conversion process, the two pulses
propagates with the common group velocity $v_g$ and do not show any
trembling (jitter) motion. If the group velocity mismatch is not
negligible ($v_{g2} \neq v_{g3}$), the solution to Eqs.(7) and (8)
is more involved, and is given by Eqs.(25) and (26) discussed in the
next section. Here we anticipate that, in this regime, the
oscillatory power transfer between the two fields is generally
accompanied by an oscillatory motion of the pulse center of mass,
which is reminiscent of ZB for the free relativistic Dirac electron.
To highlight such an analogy in a formal way, it is worth
introducing the coordinates of a moving frame
\begin{equation}
\xi=z \; , \; \; \eta=t-z/v_g
\end{equation}
where the velocity $v_g$ is defined by the relation
\begin{eqnarray}
\frac{1}{v_g}=\frac{1}{2} \left(
\frac{1}{v_{g2}}+\frac{1}{v_{g3}}\right).
\end{eqnarray}
In the moving frame, Eqs.(7) and (8) take the form
\begin{eqnarray}
\left( \frac{\partial}{\partial \xi}+ \delta
\frac{\partial}{\partial \eta} \right) A_2 & = & -i \kappa A_3 \\
\left( \frac{\partial}{\partial \xi}- \delta
\frac{\partial}{\partial \eta} \right) A_3 & = & -i \kappa A_2
\end{eqnarray}
where we have set
\begin{equation}
\delta=\frac{1}{2} \left( \frac{1}{v_{g2}}-\frac{1}{v_{g3}}\right).
\end{equation}
Equations (13) and (14) are supplemented with the boundary
conditions
\begin{equation}
A_2(0, \eta)=g(\eta) \; , \; A_3(0, \eta)=0.
\end{equation}
After introduction of the spinor wave field $\psi=(A_2,A_3)^T$,
Eqs.(13) and (14) can be finally cast into the Dirac form
\begin{equation}
i \frac{\partial \psi}{\partial \xi}=-i \sigma_z \delta
\frac{\partial \psi}{\partial \eta}+\kappa \sigma_x \psi
\end{equation}
where $\sigma_x$ and $\sigma_z$ are the Pauli matrices. Note that,
after the formal change
\begin{eqnarray}
\delta & \rightarrow  & c \nonumber \\
\kappa & \rightarrow  & \frac{m c^2}{\hbar} \\
 \xi & \rightarrow & t \; ,\;  \eta  \rightarrow  x \nonumber
\end{eqnarray}
Eq.(17) corresponds to the one-dimensional Dirac equation for a
relativistic particle of mass $m$ in absence of external fields,
moving alon the $x$ axis, written in the Weyl representation
\cite{Grenier}. Therefore, the {\it temporal} evolution of the
spinor wave function $\psi$ for the Dirac particle is mapped into
the {\it spatial} evolution of the envelopes $A_2$ and $A_3$ for
signal and sum-frequency pulses, respectively, whereas the spatial
coordinate of the Dirac particle is mapped into the retarded time
$\eta$ of the optical pulses.
\begin{figure*}
\includegraphics[scale=1.3]{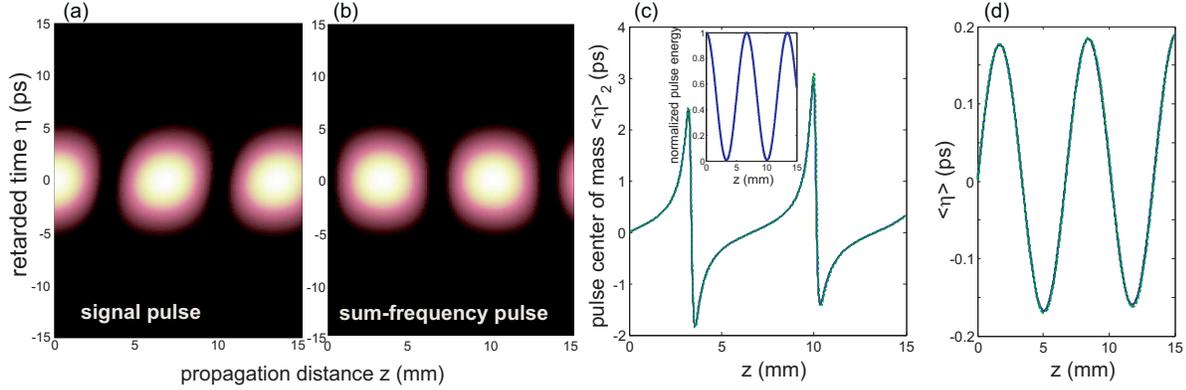}
\caption{(color online) Sum frequency generation in a
$L=1.5$-cm-long PPLN crystal showing ZB of the signal pulse. (a) and
(b): snapshots of the intensity distributions of signal [Fig.1(a)]
and sum-frequency [Fig.1(b)] pulses versus propagation distance $z$.
(c) Numerically-computed behavior of pulse center of mass $\langle
\eta \rangle_2$ for the signal field versus propagation distance
(solid curve), and corresponding behavior predicted by Eq.(28)
(dotted curve, almost overlapped with the solid one). The inset
depicts the behavior of the normalized photon fluence
$\phi_2(z)/\phi_2(0)$ of signal field versus propagation distance,
showing the oscillatory exchange of power between the signal and
sum-frequency waves. (d) Numerically-computed behavior of $\langle
\eta \rangle$, defined by Eq.(19), versus propagation distance
(solid curve), and corresponding behavior predicted by Eq.(29)
(dotted curve, almost overlapped with the solid one). The input
pulse duration is $\tau_p=5$ ps. Other parameter values are given in
the text.}
\end{figure*}
\begin{figure*}
\includegraphics[scale=1.2]{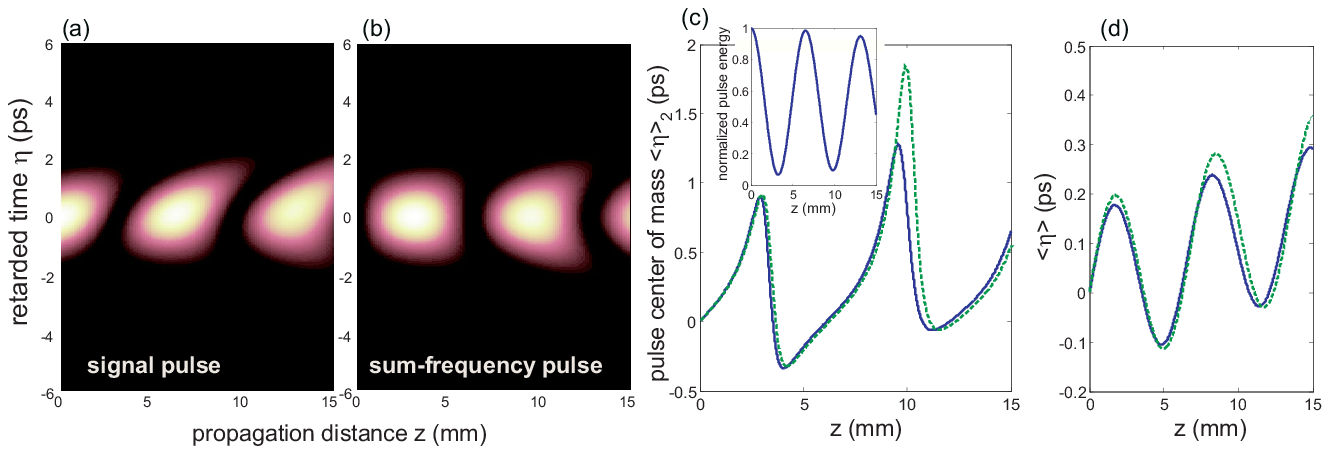}
\caption{(color online) Same as Fig.1, but for an input pulse
duration $\tau_p=1.5$ ps.}
\end{figure*}
\begin{figure*}
\includegraphics[scale=1.08]{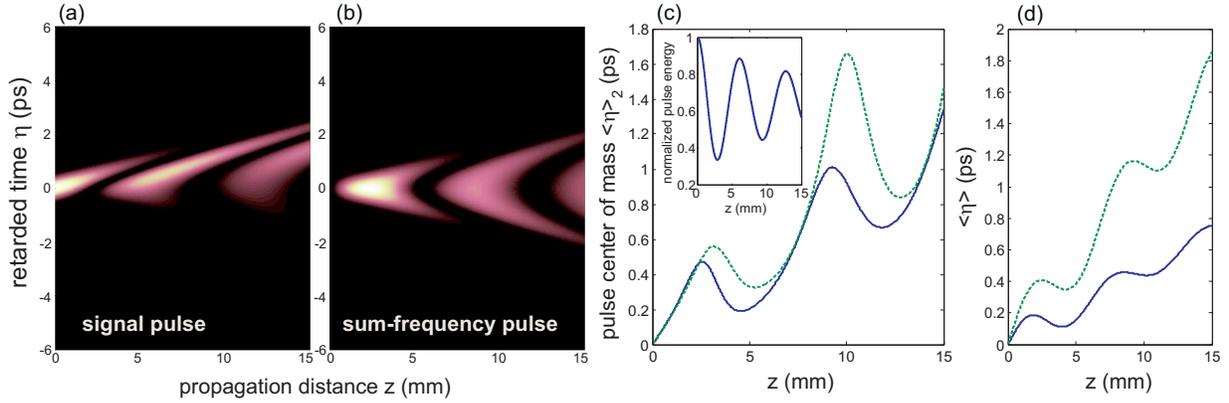}
\caption{(color online) Same as Fig.1, but for an input pulse
duration $\tau_p=0.5$ ps.}
\end{figure*}

\section{Zitterbewegung of optical pulses}
For the Dirac equation (17), ZB refers to the rapid oscillatory
motion of the expectation value of the particle position
\begin{equation}
\langle \eta \rangle (\xi) \equiv \frac{\int_{-\infty}^{\infty} d
\eta \; \eta \left( |A_2|^2+|A_3|^2 \right)}{\int_{-\infty}^{\infty}
d \eta \left( |A_2|^2+|A_3|^2 \right)}
\end{equation}
around its mean trajectory, which arises whenever negative- and
positive-energy eigenstates of the Dirac equation are simultaneously
excitated by the initial condition. Note that, indicating by
$\langle \eta \rangle_2(\xi)$ and $\langle \eta \rangle_3(\xi)$ the
(temporal) center of mass of the signal and sum-frequency pulses at
the crystal plane $\xi=z$, i.e.
\begin{equation}
\langle \eta \rangle_{2,3} (\xi) \equiv
\frac{\int_{-\infty}^{\infty} d \eta \; \eta |A_{2,3}(\xi,
\eta)|^2}{\int_{-\infty}^{\infty} d \eta |A_{2,3}(\xi,\eta)|^2}
\end{equation}
one can write
\begin{equation}
\langle \eta \rangle (\xi)=\frac{\phi_2(\xi) \langle \eta
\rangle_{2} (\xi)+\phi_3(\xi) \langle \eta \rangle_{3}
(\xi)}{\mathcal{I}_2}
\end{equation}
where $\phi_{2,3}(\xi)=\int d \eta |A_{2,3}(\xi,\eta)|^2$ are the
photon fluences of the signal and sum-frequency pulses at the plane
$z=\xi$, respectively, and
$\mathcal{I}_2=\phi_2(\xi)+\phi_3(\xi)=\phi_2(0)$ is the Manley-Rowe
invariant. In particular, if the initial pulse envelope $g(t)$ is
symmetric, i.e. $g(-t)=g(t)$, as it will be shown below one has
$\langle \eta \rangle_{3} (\xi)=0$, and thus according to Eq.(21)
the ZB of the Dirac particle for Eq.(17) can be simply retrieved
from the temporal jitter $\langle \eta \rangle_{2}(\xi)$ and
fractional energy $\phi_2(\xi)/ \phi_2(0)$ of the
signal pulse.\\
The solution to Eqs.(13) and (14) with $\delta \neq 0$ and with the
boundary conditions (16) can be readily obtained in the spectral
(Fourier) domain. Indicating by $\hat{A}_{2,3}(\xi,\omega)=(1/2 \pi)
\int d \eta A_{2,3}(\xi, \eta) \exp(-i \eta \omega)$ the spectra of
the signal and sum-frequency fields at the propagation plane $\xi$,
one has
\begin{eqnarray}
\hat{A}_{2}(\xi,\omega) & = & \hat{g}(\omega) \left[ \cos(\beta
\xi)-i \frac{\omega \delta}{\beta} \sin (\beta \xi) \right] \\
\hat{A}_{3}(\xi,\omega) & = & - \frac{i \kappa}{\beta}
\hat{g}(\omega) \sin (\beta \xi)
\end{eqnarray}
where $\hat{g}(\omega)=(1/2 \pi) \int d \eta g(\eta) \exp(-i \eta
\omega)$ is the spectrum of the signal pulse incident onto the
crystal at $\xi=0$, and
\begin{equation}
\beta(\omega)=\sqrt{\kappa^2+\omega^2 \delta^2}.
\end{equation}
In the temporal domain, the inverse Fourier transform of Eqs.(22)
and (23) yields the following exact solution for the sum-frequency
and signal pulses traveling along the crystal
\begin{equation}
A_3(\xi,\eta)=-\frac{i \kappa}{2 \delta} \int_{-\delta \xi}^{\delta
\xi} d \theta g(\theta+\eta) J_0 \left( \kappa
\sqrt{\xi^2-\frac{\theta^2}{\delta^2}} \right)
\end{equation}
\begin{eqnarray}
A_2(\xi,\eta) & = & \frac{g(\eta+\delta \xi)+g(\eta-\delta \xi)}{2}
\nonumber \\
& + &\frac{1}{2 \delta} \int_{-\delta \xi}^{\delta \xi} d \theta
g(\theta+\eta) \frac{\partial}{\partial \xi} J_0 \left( \kappa
\sqrt{\xi^2-\frac{\theta^2}{\delta^2}} \right) \nonumber \\
& -& \frac{1}{2} \int_{-\delta \xi}^{\delta \xi} d \theta
\frac{\partial g}{\partial \eta} (\theta+\eta) J_0 \left( \kappa
\sqrt{\xi^2-\frac{\theta^2}{\delta^2}} \right)
\end{eqnarray}
where $J_0$ is the zero-order Bessel function of first kind. To
calculate $\langle \eta \rangle$, $\langle \eta \rangle_2$ and
$\langle \eta \rangle_3$, let us assume, for the sake of simplicity,
that the signal pulse envelope $g(t)$ at the input crystal plane has
a symmetric profile (with e.g. a Gaussian or a sech shape)
satisfying the condition $g(-\eta)=g(\eta)$. In this case, from
Eq.(25) it follows that $A_3(\xi,-\eta)=A_3(\xi,\eta)$, and thus
\begin{equation}
\langle \eta \rangle_3(\xi)=0 \; , \; \; \langle \eta \rangle
(\xi)=\frac{\phi_2(\xi)}{\phi_2(0)} \langle \eta \rangle_2(\xi).
\end{equation}
The explicit expression of $\langle \eta \rangle_2(\xi)$, as
obtained by substitution of Eq.(26) into Eq.(20), turns out to be
rather cumbersome and not of easy physical interpretation. However,
for a signal spectrum $\hat{g}(\omega)$ narrow at around $\omega=0$
with a spectral width $\Delta \omega$ much smaller than $\kappa /
\delta$, i.e. for a relatively long input pulse, simple approximate
expressions for $\langle \eta \rangle_2(\xi)$ and $\langle \eta
\rangle(\xi)$ can be obtained, which read explicitly
\begin{eqnarray}
\langle \eta \rangle_2 (\xi) & \simeq & \xi
\frac{\delta^3}{\kappa^2} \frac{\Delta \omega^2}{\cos^2(\kappa
\xi)+\frac{\delta^2 \Delta \omega^2}{\kappa^2}}\nonumber \\
& + & \frac{\delta}{2 \kappa} \frac{\sin(2 \kappa
\xi)}{\cos^2(\kappa \xi)+\frac{\delta^2 \Delta \omega^2}{\kappa^2}}
\\
\langle \eta \rangle (\xi) & \simeq  & \xi \frac{\delta^3 \Delta
\omega^2}{\kappa^2}+ \frac{\delta}{2 \kappa} \sin(2 \kappa \xi)
\end{eqnarray}
where $\Delta \omega$ is the spectral width of the input signal
pulse, defined by
\begin{equation}
\Delta \omega^2=\frac{\int d \omega \omega^2
|\hat{g}(\omega)|^2}{\int d \omega |\hat{g}(\omega)|^2}.
\end{equation}
Equation (29) corresponds to the well-known approximate expression
of ZB in relativistic quantum mechanics (see, for instance,
\cite{Grenier,NatureZB}), whereas Eq.(28) shows the signature of ZB
in the oscillatory motion of the signal pulse center of mass as it
propagates along the crystal. Note that such an oscillatory motion,
that arises from the second term on the right hand side of Eq.(29),
is superimposed to a slight drift term [the first term on the right
hand side of Eq.(29)]. The oscillatory motion of $\langle \eta
\rangle$ and $\langle \eta \rangle_2$ along the propagation
coordinate $\xi$ of the crystal basically follows the
oscillatory-like of optical transfer between signal and
sum-frequency pulses. Note also that, according to Eq.(28) and
because of the assumption $\delta \Delta \omega / \kappa \ll 1$, the
pulse center of mass $\langle \eta \rangle_2(\xi$ takes large values
at the propagation distances $\xi=\pi/2 \kappa$,
 $\xi= \pi/\kappa$, $\xi=3 \pi/ 2\kappa$, ... where most of the signal field is converted into the
 sum-frequency field.\\
As an example, let us consider the process of sum frequency
generation in a nonlinear periodically-poled lithium-niobate (PPLNB)
crystal (see, for instance, \cite{Fejer}) assuming $\lambda_1=1550$
nm, $\lambda_2=810$ nm, and $\lambda_3=532$ nm for the wavelengths
(in vacuum) of pump, signal and sum-frequency waves, respectively.
From Sellmeir equations \cite{Sell}, one can estimate at 25$^{\rm
o}$C and for extraordinary waves $n_1=2.1381$, $n_2=2.1748$,
$n_3=2.2343$, $v_{g2}/c_0 = 0.4422$, $v_{g3}/c_0 = 0.4069$ and a QPM
period of $\Lambda= 2 \pi/ \Delta k \simeq
 7.39 \; \mu$m, which is easily accessible with current poling
 technology. For first-order QPM with alternating sign +/- of
 $\chi^{(2)}$ with period $\Lambda/2$, the effective nonlinear
 coefficient is given by \cite{Fejer} $d_{eff}=(2/\pi)d$, where $d$ is the element of the nonlinear $d$-tensor
of the crystal involved in the parametric interaction ($d=d_{33}
\simeq 27 \; {\rm pm/V}$ for extraordinary waves). As an input
signal pulse, we assume a Gaussian profile $g(t) \propto
\exp(-t^2/\tau_0^2)$ with a full-width at half maximum (FWHM) pulse
duration $\tau_p= (\sqrt{2 {\rm log} 2}) \tau_0$ and spectral pulse
width $\Delta \omega=1/ \tau_0$. As an example, Figs.1(a) and 1(b)
show the evolution of the pulse intensity profiles $|A_2(\xi,
\eta)|^2$ and $|A_3(\xi, \eta)|^2$ of signal and sum-frequency
fields, respectively, in a $L=$1.5-cm-long PPLN crystal
 as obtained by direct numerical analysis of
Eqs.(1-3), for a signal pulse duration $\tau_p=5$ ps of low peak
intensity ($1$ W/cm$^2$) and an intensity of the continuous-wave
pump field $I_1=\hbar \omega_1 |A_1|^2$ of $1 \; {\rm MW/cm^2}$,
corresponding to $\kappa \simeq 0.4656 \; {\rm mm}^{-1}$ and $
\delta \Delta \omega / \kappa \simeq 0.0827$. The numerical results
of Fig.1 are with excellent accuracy reproduced by the analytical
solutions (25) and (26), derived in the no-pump depletion limit.
Figures 1(c) and 1(d) show the corresponding behavior, along the
crystal coordinate $\xi=z$, of the normalized photon fluence
$\phi_2(\xi)/ \phi_2(0)$ [inset of Fig.1(c)], pulse center of mass
$\langle \eta \rangle_2(\xi)$ of signal field [solid curve in
Fig.1(c)], and ZB amplitude $\langle \eta \rangle (\xi)=[\phi_2(\xi)
/ \phi_2(0)] \langle \eta \rangle_2 (\xi)$ [solid curve in
Fig.1(d)]. In Figs.1(c) and (d), the behaviors of $\langle \eta
\rangle_2 (\xi)$ and $\langle \eta \rangle (\xi)$, as predicted by
Eqs.(28) and (29), are also shown (dotted curves, almost overlapped
with the solid ones). Note that, according to the theoretical
analysis, the center of mass of the signal pulse undergoes a clear
oscillatory motion, superimposed to a slight drift [arising from the
first term on the right hand side of Eq.(28)]. For spectrally
broader signal pulses, ZB can not be accurately described by the
simple Eqs.(28) and (29), however the oscillatory motion of the
pulse center of mass, superimposed to a drift motion, is still
observed in numerical simulations. This is shown, as an example, in
Figs.2 and 3, where pulse duration of the injected signal pulse has
been reduced to $\tau_p=1.5$ ps in Fig.2 (corresponding to $ \delta
\Delta \omega / \kappa \simeq 0.2758$), and to $\tau_p=0.5$ ps in
Fig.3 (corresponding to $ \delta \Delta \omega / \kappa \simeq
0.827$). In an experiment, a measurement of the pulse center of mass
at internal planes of the nonlinear crystal could be a nontrivial
task, whereas autocorrelation measurements can readily reveal a
jitter of the output pulse, at the exit of the crystal, with respect
to a reference case. Owing to the dependence of the sinusoidal terms
entering in Eq.(28) on the product $\kappa \xi \propto \sqrt{I_1}
\xi$, in an experiment the signature of ZB can be easier revealed by
monitoring the center of mass of the signal pulse at the output
plane $\xi=L$ of the crystal as a function of the pump intensity
$I_1$. This is shown, as an example, in Fig.4, where the
numerically-computed behavior of $\langle \eta \rangle_2$ at the
output crystal plane versus the intensity $I_1$ of the pump field is
depicted, together with the approximate prediction based on
Eq.(28).\\

\begin{figure}
\includegraphics[scale=1]{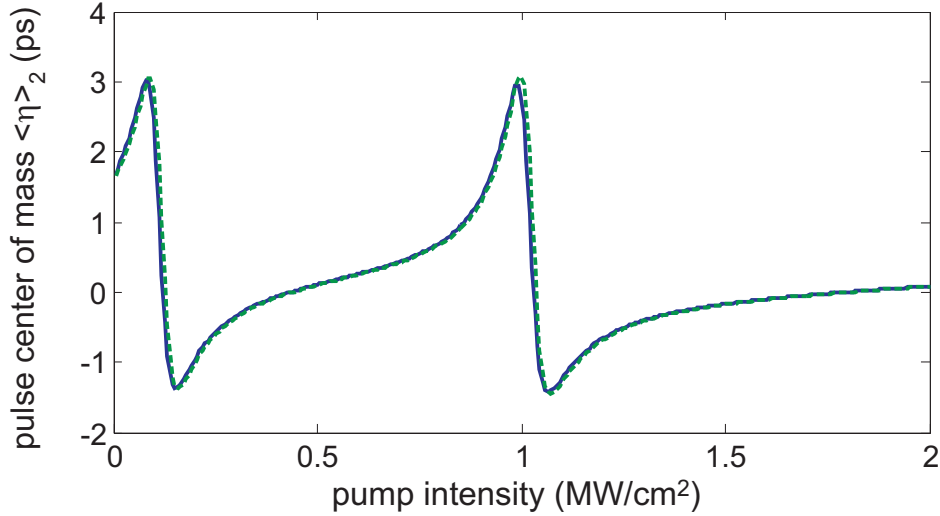}
\caption{(color online) Numerically-computed behavior of the center
of mass $\langle \eta \rangle_2$ for the signal pulse at the output
of the PPLN crystal versus the intensity $I_1$ of the pump field.
The crystal length is $L=1$ cm, the input pulse duration of the
signal field is $\tau_p=5$ ps. The dotted curve in the figure is the
behavior of $\langle \eta \rangle_2$ predicted by Eq.(28).}
\end{figure}

\section{Conclusions}
In this work, a photonic analogue of the trembling motion
(Zitterbewegung) of a free relativistic Dirac particle, based on
frequency conversion of short optical pulses in a nonlinear
quadratic medium, has been presented. The analogy, which stems from
the mathematical similarity between the Dirac equation of a massive
particle and the coupled equations describing sum frequency
generation in the presence of pulse walk-off, indicates that
well-known and experimentally accessible nonlinear optical processes
could be exploited to simulate the Dirac equation in an optical
setting. As compared to other classical and quantum analogues of
Zitterbewegung recently proposed in the literature, based on trapped
ions \cite{ZB2,NatureZB}, graphene \cite{ZB3,ZB4_1,ZB4-2,ZB4-3} or
photonic crystals, superlattices or metamaterials \cite{P1,P2,P3},
our proposal may show a simpler experimental access and could
stimulate further search for nonlinear optics analogues of
relativistic quantum phenomena. For example, engineering of the QPM
grating could be exploited to introduce in Eq.(17) a
$\xi$-dependence of $\kappa$, i.e. to simulate the dynamics of a
relativistic Dirac particle with a time-varying mass
\cite{timevarying}. Likewise, if the temporal dependence of the pump
pulse is included in the analysis and assuming $v_{g1}=v_g$, an
$\eta$-dependence of the mass $\kappa$ is introduced in Eq.(17),
which enables to mimic a relativistic Dirac particle in a Lorentz
scalar potential \cite{Grenier}.\\
\\
 The author acknowledges financial support by the
Italian MIUR (Grant No. PRIN-2008-YCAAK project "Analogie
ottico-quantistiche in
strutture fotoniche a guida d'onda").\\

\end{document}